\documentclass[aps,prl,reprint]{revtex4-2}

\usepackage{graphicx}
\usepackage{xcolor}
\usepackage{siunitx}

\begin{document}

\title{Deep neural networks for the prediction of the optical properties\\
       and the free-form inverse design of metamaterials}

\author{Timo Gahlmann}
 \affiliation{Department of Physics, Chalmers University of Technology, SE-41296 G\"oteborg, Sweden}
\author{Philippe Tassin}
 \affiliation{Department of Physics, Chalmers University of Technology, SE-41296 G\"oteborg, Sweden}

\date{\today}

\begin{abstract}
Many phenomena in physics, including light, water waves, and sound, are described by wave equations. Given their coefficients, wave equations can be solved to high accuracy, but the presence of the wavelength scale often leads to large computer simulations for anything beyond the simplest geometries. The inverse problem, determining the coefficients from a field on a boundary, is even more demanding, since traditional optimization requires a large number of forward problems be solved sequentially. Here we show that the free-form inverse problem of wave equations can be solved with machine learning. First we show that deep neural networks can be used to predict the optical properties of nanostructured materials such as metasurfaces. Then we demonstrate the free-form inverse design of such nanostructures and show that constraints imposed by experimental feasibility can be taken into account. Our neural networks promise automated design in several technologies based on the wave equation.
\end{abstract}

\maketitle

Machine learning is a technology that has revolutionized medical diagnosis, image and face recognition, self-driving vehicles, and many more applications in the past decades. Recently, neural networks have also led to new developments in physics. For example, it was shown that neural networks can find phase transitions in quantum-mechanical systems~\cite{VanNieuwenburg}, predict the ultrafast dynamics in nonlinear fibre
optics~\cite{Salmela}, and even identify basic physical concepts such as energy conservation~\cite{Iten}. The laws of physics are often described by partial differential equations and a formidable number of computational techniques have been developed to solve partial differential equations given definite values of their coefficients. The inverse problem~\cite{Whiting,Kalt,Alcalde,Pestourie,Hegde,Kudyshev}, determining the coefficients that produce an a priori given solution, is a much more complex problem, but a very relevant one since it may offer technological solutions that go beyond the human imagination. Free-form inverse design in physics, also known as topology optimization, is well established in continuum mechanics~\cite{Bensoe}, but only at its infancy in paradigms of physics that rely on the wave equation~\cite{Jensen}.

In this article, we will show that the problem of free-form inverse design for wave equations can be solved with deep neural networks. To solve the inverse problem of the wave equation, one normally defines an objective function that characterizes how close the solution of the wave equation is to the desired solution. An optimization algorithm is then used to find the coefficients of the wave equation that minimize the objective function. This can be applied to the design of, for example, optical and acoustic devices, since the coefficients of the wave equation define the materials and the geometry of the device. Depending on the number of degrees of freedom, inverse design can be achieved in different ways. With three or fewer parameters, an exhaustive parameter scan of the wave equation's solution can often be obtained in a reasonable time. If the wave equation's coefficients must be represented by three to a few tens of parameters, binary search~\cite{BingShen}, physics-informed gradient-descent methods~\cite{LalauKeraly,Piggott,Beilina,Michaels,Black}, and stochastic optimization such as genetic algorithms~\cite{Elsawy,Molesky,Jensen,Campbell_2019,Osher} and particle swarm optimization~\cite{Preble,Shokooh,Wiecha,Liu}, have been shown effective. Such optimization algorithms require a large number of evaluations of the forward problem, which must be solved sequentially. Sometimes, inverse design reproduces known structures such as periodic photonic crystals or gratings~\cite{Beilina,Piggott2020}, but more often capricious designs with small features that are difficult to fabricate with lithographic methods are obtained; this shows the importance of integrating  fabrication feasibility into the inverse design method.

\begin{figure*}
	\centering
	\includegraphics[width=1.0\textwidth]{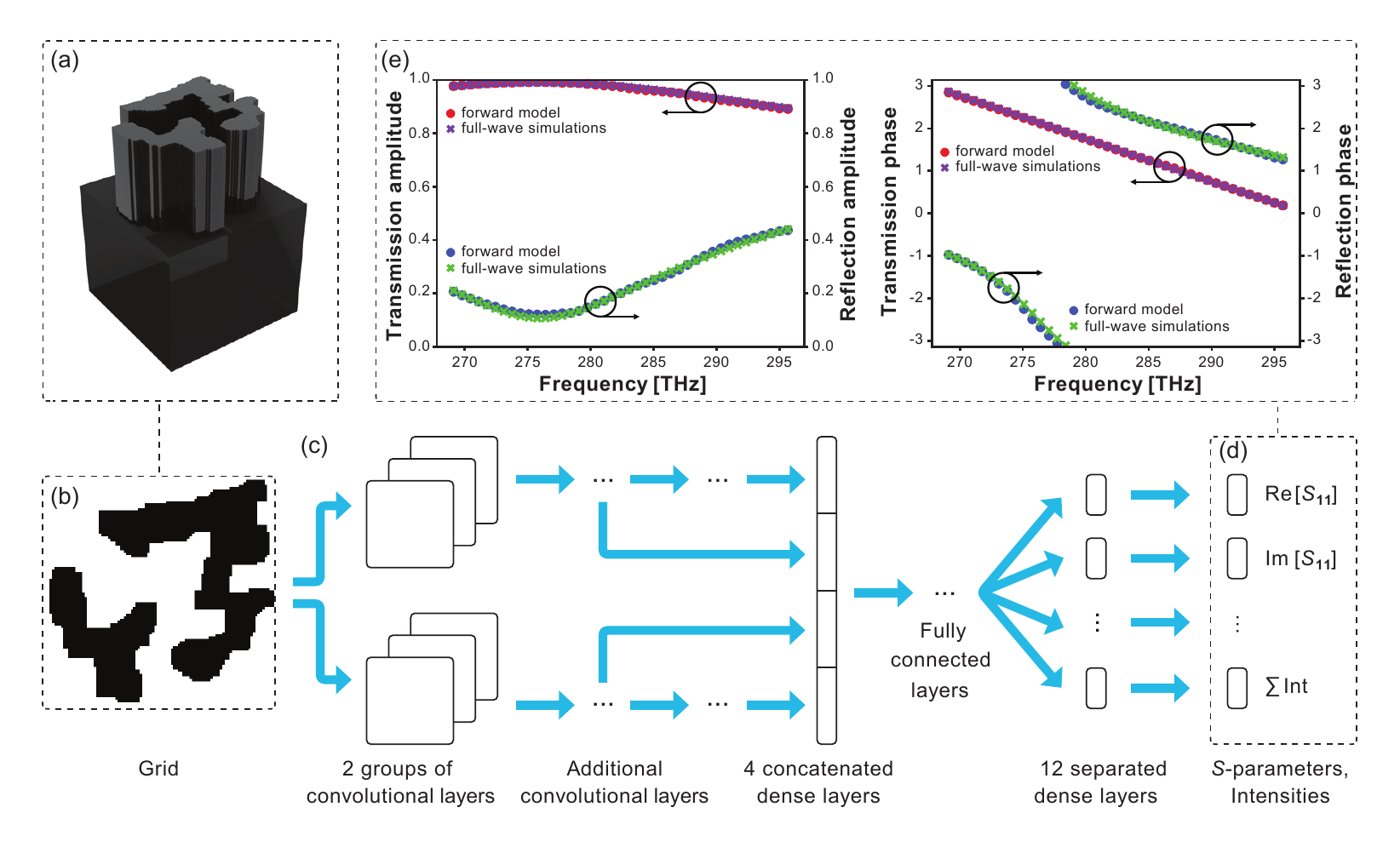}
	\caption{The forward model, which is schematically depicted in this figure, approximates the $\mathcal{S}$ parameters of a given metasurface meta-atom. (a) Rendered illustration of a meta-atom. (b) Grid representation (2D cross section) of the meta-atom. (c) Schematic architecture of the forward model. (d) Output of the forward model. (e) Comparison of $\mathcal{S}$ parameters predicted by the forward model (circles) and calculated by full-wave simulations (crosses).}
	\label{Fig:Forward}
\end{figure*}

Here we want to specifically focus on metasurfaces, i.e., thin nanotechnological structures that are built up from subwavelength-small elementary building blocks---also called meta-atoms~\cite{Genevet,Ding,Kamali}. These meta-atoms are individually designed to locally modify the phase and amplitude of the scattered waves, allowing us to control the electromagnetic wave front at a subwavelength scale. In this way, one can obtain nontrivial scattering response from the metasurface. Metasurfaces have been designed for a wide range of different purposes, e.g., for holograms~\cite{GZheng}, axicons~\cite{DLin}, and retroreflectors~\cite{Arbabi}. A metasurface consists of a two-dimensional array of thousands of meta-atoms, each of which needs to be individually designed to have the correct transmission amplitude and phase. With a gradient-descent or stochastic optimization method, this means the optimization must be restarted thousands of times, leading to wall-clock times of months or years. Below we will discuss how it is possible to address the computational and methodological difficulties emerging when using  neural networks for the inverse design of metasurfaces~\cite{Khoram,J.Mun,L.Gao,F.Cheng_2019,L.Gao,X.Shi}. Indeed, deep artificial neural networks (DNNs) can have significant advantages compared to traditional calculation methods when they are applied carefully~\cite{Goodfellow_2016,LeCun,Wiecha-review}. In particular, DNNs provide the big advantage that they can quickly generate a design for a meta-atom with any desired transmission amplitude and phase in the subset of realizable $\mathcal{S}$ parameters once they are trained, with no need to run an optimization for each meta-atom again. The training of the neural networks still requires a large number of simulations of the wave equation, but these simulations are independent, which means they can be run simultaneously on high-performance computer clusters.

We start with building a forward model $\mathcal{F}$ that approximates the optical properties~\cite{Malkiel,Fowler,Campbell_2020} of our meta-atoms, the scattering parameters $\mathcal{S}$, as a function of the grid $(g)$ representing the lithographic mask:
\begin{equation}
    \mathcal{F}[g]\approx \mathcal{S}
\end{equation}
Such a forward model can considerably speed up the evaluation of the optical properties compared to direct full-wave solutions of the wave equation. Later in this paper, we will use the forward model as a surrogate model~\cite{Zabaras,Chugh,Whiting,Pestourie} in the training of the CGAN model we will build for the inverse design; indeed, a fast and differentiable approximation of the optical properties is required for the convergence of the CGAN model to output high-quality results in a reasonable time.

\begin{figure*}[t!]
	\centering
	\includegraphics[width=1.0\textwidth]{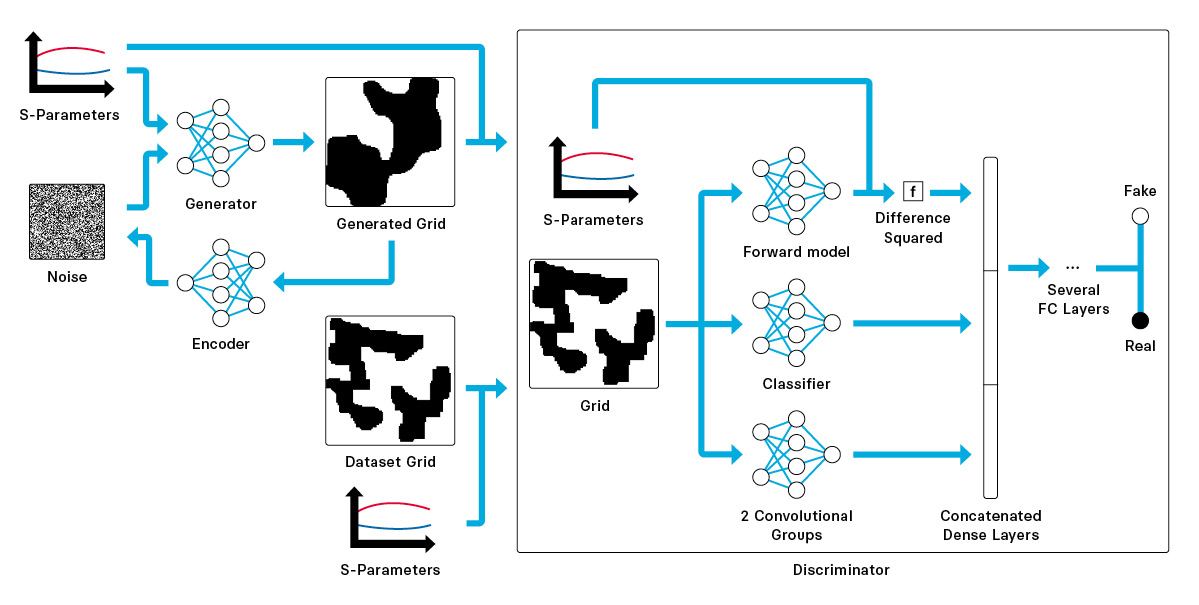}
	\caption{The conditional generative adversarial network model. The generator creates meta-atom grids from noise and $\mathcal{S}$ parameters, while the encoder prevents mode collapse. Two expert networks (forward and classifier model) are integrated inside the discriminator.}
	\label{Fig:cGAN}
\end{figure*}

Our metasurface consist of a silicon substrate with a \SI{300}{\nano\meter}-thin silica spacer and a \SI{500}{\nano\meter}-thick patterned layer of silicon. We use a binary grid to represent the lithographic mask---0 means that the top layer is etched away, 1 means the top layer remains at that position. An example of a meta-atom is shown in Fig.~\ref{Fig:Forward}(a)-(b). This prepares us for free-form inverse design where the geometry is not defined by shape parameters. Due to the similarity of this binary grid representation with pixelated images, a convolutional neural network (CNN) was chosen to model the $\mathcal{S}$ parameters of the meta-atoms. (The topology of the neural network is shown in Fig.~\ref{Fig:Forward}(c)-(d) and in Fig.~S1.) The network approximates the transmission amplitude, intensity, real and imaginary part of each of the four $\mathcal{S}$ parameters (transmission and reflection in two polarization states), and  the sum of all intensities for 41 equally spaced frequencies. These 17 values for each frequency are partly redundant, which is intended to stabilize the training and output of the CNN by injecting energy conservation and other relationships between the $\mathcal{S}$ parameters into the objective function. In order to create the training and validation data set for the CNN, about 40.000  semirandom grids were created and loaded into the finite-element simulator COMSOL~\cite{COMSOL} to obtain the $\mathcal{S}$ parameters. The binary grids were grown from random noise by filling too small holes or erasing islands of increasing size, since such holes and islands cannot be realized by current lithographic techniques. In some cases symmetry operation were applied. Therefore, the allowed geometries of our meta-atoms are only limited by the $100 \times 100$ resolution of the grid and by fabrication feasibility, and not by any other constraints arising from the computational method.

In Fig.~\ref{Fig:Forward}(e), we show the results for a sample from the validation set. We find excellent agreement between the $\mathcal{S}$ parameters predicted by the forward model and the $\mathcal{S}$ parameters calculated from the finite-element simulations (with the exception of samples with narrow resonances, see also below), with a mean square error in the validation set of less than 0.003. To achieve these results, about 36.000 samples were used for training and 4.000 for validation.

As we argued before, it is important to integrate fabrication feasibility into the inverse design method. Filtering out too small geometry features as we did for the creation of the training samples is, however, too time-consuming. To address this issue, we trained a classifier network $\mathcal{C}$ on flawless and flawed grids:
\begin{equation}
    \mathcal{C}[g]\approx \mathcal{D}
\end{equation}
The grids were labeled with the number of pixels $(\mathcal{D})$ that would need to be changed in order to satisfy the fabrication restrictions by a variation of the algorithm that created the semi-random grids for the training samples. The complete layout of the classifier network can be found in Fig.~S2.

We now have the elements we need to build the neural network model for the inverse design. The traditional methods of using an artificial neural network (ANN) as a surrogate model for evolutionary optimization~\cite{Hegde} are computationally not the most efficient and often bare the risk of converging to a singular point of the surrogate model. An exclusively ANN-based inverse design model is much more computationally efficient and allows to obtain multiple possible solutions for one set of target scattering parameters~\cite{Malkiel,Fowler,Zhelyeznyakov}. The main obstacle for the convergence of an inverse design network is the so-called ``one-to-many'' problem, which describes the issue that in general multiple non-unique solutions exist for a given design target $(\mathcal{F}[g_i]-\mathcal{F}[g_j]<\epsilon)$. In consequence, a naive inversion of the ANN fails to converge, because the network is trained on seemingly conflicting data, where one set of input values has different possible output targets ($\mathcal{F}^{-1}[\mathcal{S}]=g_i$ and $\mathcal{F}^{-1}[\mathcal{S}]=g_j$). Therefore, the network cannot converge to one of the local minima of the loss function. One way of dealing with the one-to-many problem is a so-called conditional generative adversarial network (CGAN)~\cite{B.Zheng,J.Jiang,P.Rodrigues,S.SoandJ.Rho,A.Mall}, which is a variation of the traditional GAN~\cite{Goodfellow_2020} (see Fig.~\ref{Fig:cGAN} for a schematic of the topology of the CGAN network and Fig.~S3 and Fig.~S4 for the full network topology). In a CGAN, a generator $\mathcal{G}$ creates grids based on noise $\mathcal{N}$ and the desired $\mathcal{S}$ parameters, while the discriminator $\mathcal{D}$ is fed not only the grids produced by the generator and grids from the given data set, but also the respective labels ($\mathcal{S}$ parameters):
\begin{equation}
    \mathcal{G}[\mathcal{S},\mathcal{N}]=g_{\mathcal{G}} \quad \mathcal{D}[g,\mathcal{S}]=o_{\mathcal{D}}
\end{equation}
The discriminator takes the grids as input and has in this case two groups of convolutional layers that analyse them, where one of the two groups is a residual network~\cite{ResNet}. Combined with subsequent fully connected layers, this should in theory be sufficient to train the discriminator. However, we have found that it is a difficult task for the discriminator to differentiate between grids produced by the generator and grids from the data set while at the same time evaluating whether the grids have the desired optical properties and obey the required fabrication constraints. Therefore, we have integrated two pre-trained expert networks, the forward network $\mathcal{F}$ and the classifier network $\mathcal{C}$, into the discriminator. This significantly improves the training speed and performance of the discriminator and, thereby, also the performance and results of the generator:
\begin{equation}
    \mathcal{D}[g,f(\mathcal{F}(g),\mathcal{S}),\mathcal{C}(g)]=o_{\mathcal{D}}
\end{equation}
The two expert models are embedded into the discriminator by feeding the grid $g$ to the two networks and their output to a hidden layer of the discriminator after performing some simple mathematical operations, such as evaluating the difference between the $\mathcal{S}$ parameter labels and the output of the forward network and subsequent multiplication with the optional input size mask, which implements the option of only specifying a subset of desired $\mathcal{S}$ parameters.

The architecture of the generator differs from traditional ones~\cite{Goodfellow_2020}; instead of having one group with multiple batch normalizations we use 4 groups of transpose convolutional layers with varying amounts of batch normalisations in each group, which we found to work better. The four groups are stacked at the end and a single convolutional filter transforms the concatenated filters from all groups to the output of the generator. It turns out that, like GANs in general, also our CGAN with continuous labels is susceptible to mode collapse. This means that the output of the generator depends very little or not at all on the input noise and the generator maps only to a small subset of the surrogate model. Apart from the typical reasons for mode collapse, this is further enhanced by the continuous nature of the labels, which already demands a relatively large variety of the generated grids, so the discriminator cannot punish this dynamic sufficiently. Therefore, an encoder network $\mathcal{E}$ was introduced, which is trained along with the generator to map back from the generated grids and the corresponding labels to the noise input of the generator:
\begin{equation}
    \mathcal{E}[\mathcal{G}(\mathcal{S},N),\mathcal{S}]=\mathcal{E}[g_{\mathcal{G}},\mathcal{S}]\approx \mathcal{N}
\end{equation}
The resulting error of the encoder is added to the loss function of the generator. In this way, the generator is incentivized to encode the noise in the grids, so the encoder can map it back and the result is a greater variety of generated grids. Another improvement in the learning algorithm was achieved by integrating grids created in previous epochs of the same training cycle into the current epoch to stabilize the discriminator and, consequently, also the generator. Furthermore, the generator was asked to created grids with specific $\mathcal{S}$ parameters that would be most relevant after training in order to improve the quality of these generated grids. The relative share of grids created during the current or previous epochs and of asking the generator to create specific grids of interest can be preset or adaptive based on the average output of the discriminator in the current epoch.

Finally, we have iteratively improved the training data by using the CGAN to find training samples in sparsely sampled regions of the S-parameter space. This increases the accuracy of all models, reduces the number of corrections needed to be made to the grids generated by the generator in order to satisfy feasibility constraints, and leads to the $\mathcal{S}$ parameters of the generated grids better approximating the target parameters.

\begin{figure*}[t!]
	\centering
	\includegraphics[width=1.0\textwidth]{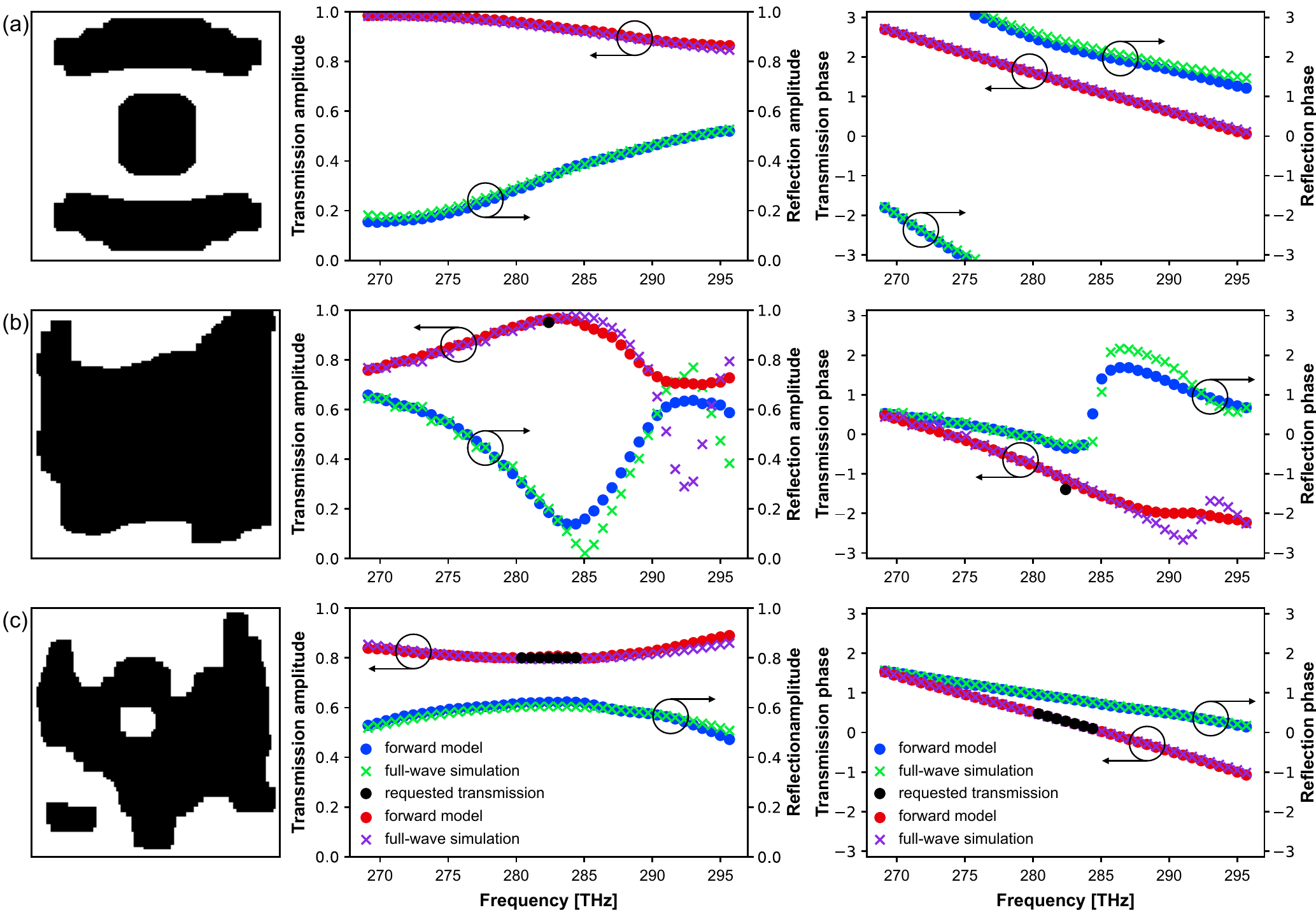}
	\caption{Comparison of the $\mathcal{S}$ parameters as requested from the CGAN (black dots), predicted by the forward model (blue and red dots), and calculated by full-wave simulations (green and purple crosses). (a) A meta-atom from the validation set. (b) A meta-atom with a resonance obtained by requesting a desired transmission amplitude and phase at a single frequency from the CGAN. (c) A meta-atom obtained from the CGAN by requesting a constant transmission amplitude and linear transmission phase.}
	\label{Fig:cGANResults}
\end{figure*}

We now demonstrate that our CGAN network is very efficient in designing meta-atoms with desired $\mathcal{S}$ parameters. In Fig.~\ref{Fig:cGANResults}, we show the transmission amplitude and phases for three meta-atoms: one meta-atom from the validation set and two meta-atoms from the generator. For each meta-atom, we show the requested $\mathcal{S}$ parameters as well as the spectrum of the $\mathcal{S}$ parameters obtained from a direct finite-element simulation and the output of the forward model. We see that the agreement between the $\mathcal{S}$ parameters predicted by the CGAN and the exact calculation by full-wave simulations is excellent, not only for meta-atoms from the validation set [Fig.~\ref{Fig:cGANResults}(a)], but also for meta-atoms created by the generator [Fig.~\ref{Fig:cGANResults}(b)-(c)]. Resonances are not captured perfectly by the CGAN and forward model, but this is not important for our purposes, since we are most often interested in meta-atoms with low frequency dispersion. Meta-atoms with resonances can still be used by the CGAN and are even important to reach certain transmission phases [black dot in Fig.~\ref{Fig:cGANResults}(b)], even if this phase is probed at a frequency outside the linewidth of the resonance. It is equally possible to request a desired spectrum from the CGAN. In Fig.~\ref{Fig:cGANResults}(c), for example, we request a spectrum with constant amplitude and linear phase, and again we get excellent agreement between the requested $\mathcal{S}$ parameters and full wave simulations.

\begin{figure*}[t!]
	\centering
	\includegraphics[width=\textwidth]{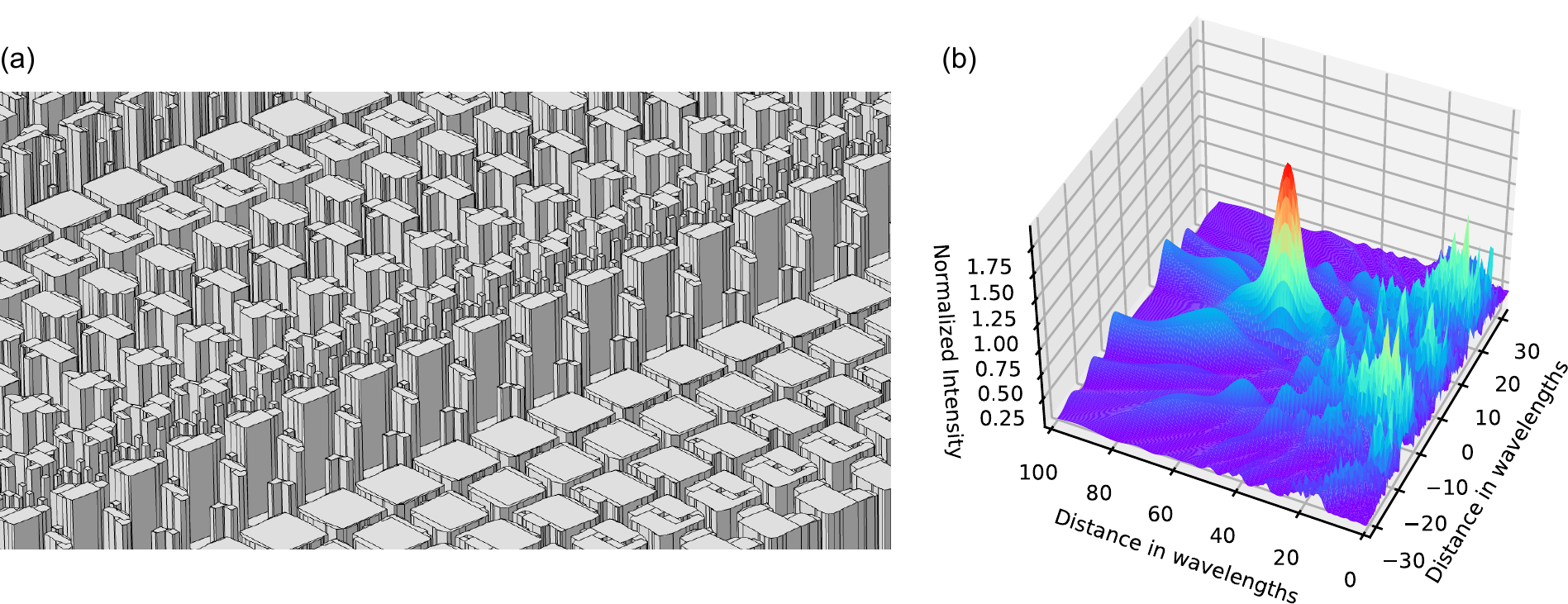}
	\caption{Free-form inverse design of a metasurface cylindrical lens. (a)~The array of meta-atoms with predetermined phase generated by the CGAN that constitute the lens. (b)~Optical intensity of the beam focused by the ANN-designed lens. The designed focal distance was 50 wavelengths away from the metasurface.}
	\label{Fig:cyl}
\end{figure*}

Finally, we have designed a cylindrical metasurface lens. A lens is essentially a phase mask that focuses an incoming plane wave to a focal point. This can be achieved in a metasurface by stacking meta-atoms in a two-dimensional array where each meta-atom is designed to have a predetermined phase (and constant amplitude). Figure~\ref{Fig:cyl}(a) shows the central section of such a metasurface lens designed with our CGAN. Indeed, the CGAN allows us to obtain very quickly a set of meta-atoms with any transmission phase between 0 and $2\pi$ and with an almost constant amplitude. Constructing such a map without DNN would have required an exhaustive parameter search and, to the best of our knowledge, this would not have led to meta-atoms with constant amplitude. Subsequently, we perform a full-wave simulation of how a normally-incident optical beam is focused by our metasurface. Plotting the optical intensity around the designed focal length of 50 wavelengths away from the metasurface in Fig.~\ref{Fig:cyl}(b), we observe a sharp peak in intensity confirming that the metasurface lens is focusing light exactly onto the desired focal point. This confirms the successful inverse design of the wave equation by our CGAN.

\section{Conclusion}
We have applied machine learning to the problem of inverse design of nanophotonic devices. We have found that a conditional generative adversial network together with a classifier for fabrication feasibility constraints and a surrogate neural network model for predicting the optical properties is particularly suited for inverse design in nanophotonics. Our trained neural network is able to generate thousands of designs for meta-atoms in seconds, allowing us to design large-area metasurfaces with arbitrary amplitude and phase masks (in the range of parameters provided in the training set)---all meta-atoms in such a metasurfaces can be generated without needing to run an optimization algorithm for every meta-atom individually. The network also has fabrication feasibility incorporated, avoiding the capricious designs that are obtained with some other methods. The training of the neural networks requires a large number of full-wave simulations obtained with a classical solver for partial differential equations. However, a big advantage of deep neural networks over iterative optimization methods is that these simulations are independent and can be run simultaneously on different computers, allowing us to take advantage of high-performance cluster infrastructure. We expect that our conditional generative adversial network, particularly in combination with dimensionality reduction techniques~\cite{Adibi}, can become a widely adopted tool to achieve inverse design in nanophotonic structures, such as metasurfaces, metamaterials, plasmonic waveguides and resonators, and nanophotonic structures for nonlinear optics. While we have focused in this article on metasurfaces with a certain desired scattering behaviour, our network can be employed equally easily to design structures with strong near fields or large nonlinear optical interactions. Finally, since our training samples are obtained from full-wave simulations of the wave equation, our approach can be generalized to other technology areas based on the wave equation, such as microphotonics, acoustics, water waves, and seismic waves.

\section{Acknowledgement}
We acknowledge support from Chalmers’ Excellence Initiative Nano and from the Swedish Research Council under Grant
No.~2020-05284. The numerical calculations and ANN training were performed on resources provided by the Swedish
National Infrastructure for Computing (SNIC) at the Chalmers Centre for Computational Science and Engineering (C3SE), partially funded by the Swedish Research Council under Grant No. 2018-05973.


\begin{thebibliography}{1}
   \bibitem{VanNieuwenburg}
    E. P. L. {van} Nieuwenburg, Y.-H. Liu, and S. D. Huber, “Learning phase transitions by confusion,” Nat. Physics 13, 435–439 (2017).
   \bibitem{Salmela}
    L. Salmela , N. Tsipinakis, A. Foi, C. Billet, J. M. Dudley, and
    G. Genty, “Predicting ultrafast nonlinear dynamics in fibre optics with a recurrent neural network,” Nat. Mach. Intell.  3, 344–354 (2021).
    \bibitem{Iten}
    R. Iten, T. Metger, H. Wilming, L. del Rio, and R. Renner, “Discovering Physical Concepts with Neural Networks,” Phys. Rev. Lett. 124, 010508 (2020).
    \bibitem{Whiting}
    S. D. Campbell, D. Z. Zhu, E. B. Whiting, J. Nagar, D. H. Werner, and P. L. Werner, “Advanced multi-objective and surrogate-assisted optimization of topologically diverse metasurface architectures,” in Metamaterials, Metadevices, and Metasystems 2018, vol. 10719 (SPIE, 2018), p. 107190U.
    \bibitem{Kalt}
    V. Kalt, A. K. González-Alcalde, S. Es-Saidi, R. Salas-Montiel, S. Blaize, and D. Macías, “Metamodeling of high-contrast-index gratings for color reproduction,” J. Opt. Soc. A 36, 79–88 (2019).
    \bibitem{Hegde}
    R. S. Hegde, “Photonics Inverse Design: Pairing Deep Neural Networks With Evolutionary Algorithms,” IEEE J. Sel. Top. Quantum Electron. 26 ,1–8 (2020).
    \bibitem{Pestourie}
    R. Pestourie, Y. Mroueh, T. V. Nguyen, P. Das, and S. G. Johnson, “Active learning of deep surrogates for PDEs: Application to metasurface design,” arXiv:2008.12649 [physics] (2020).
    \bibitem{Alcalde}
    A. K. González-Alcalde, R. Salas-Montiel, V. Kalt, S. Blaize, and D. Macías, “Engineering colors in all-dielectric metasurfaces: Meta-modeling approach,” Opt. Lett. 45, 89–92 (2020).
    \bibitem{Kudyshev}
    Z. A. Kudyshev, A. V. Kildishev, V. M. Shalaev, and A. Boltasseva, “Machine learning assisted global optimization of photonic devices,” arXiv:2007.02205 [physics] (2020).
    \bibitem{Bensoe}
    M. P. Bendsøe and O. Sigmund, “Topology Optimization: Theory, Methods and Applications” (Springer, New York, 2003).
    \bibitem{Jensen}
    J. S. Jensen and O. Sigmund, “Topology optimization for nano-photonics,” Laser Photon. Rev. 5, 308-321 (2011).
    \bibitem{BingShen}
    B. Shen, P. Wang, R. Polson, and R. Menon, “An integrated-nanophotonics polarization beamsplitter with $2.4x2.4 \mu{m}^2$ footprint,” Nat. Photonics 9, 378–382 (2015).
    \bibitem{LalauKeraly}
    C. M. Lalau-Keraly, S. Bhargava, O. D. Miller, and E. Yablonovitch, “Adjoint shape optimization applied to electromagnetic design,” Opt. Express 21, 21693-21701 (2013).
    \bibitem{Piggott}
    A. Y. Piggott, J. Lu, K. G. Lagoudakis, J. Petykiewicz, T. M. Babinec, and J. Vučković, “Inverse design and demonstration of a compact and broadband on-chip wavelength demultiplexer,” Nat. Photonics 9, 374-377 (2015).
    \bibitem{Beilina}
    L. Beilina, L. Mpinganzima, and P. Tassin, “Adaptive optimization algorithm for the computational design of nanophotonic structures,” International Conference on Electromagnetics in Advanced Applications (ICEAA), 2016, 420-423 (2016).
    \bibitem{Michaels}
    A. Michaels and E. Yablonovitch, “Inverse design of near unity efficiency perfectly vertical grating couplers,” Opt. Express 26, 4766-4779 (2018).
    \bibitem{Black}
    L.-J. Black, Y. Wang, C. H. de Groot, A. Arbouet, and O. L. Muskens, “Optimal Polarization Conversion in Coupled Dimer Plasmonic Nanoantennas for Metasurfaces,” ACS Nano 8, 6390–6399 (2014).
    \bibitem{Campbell_2019}
    S. D. Campbell, D. Sell, R. P. Jenkins, E. B. Whiting, J. A. Fan, and D. H. Werner, “Review of numerical optimization techniques for meta-device design [Invited],” Opt. Mater. Express 9, 1842–1863 (2019).
    \bibitem{Elsawy}
    M. M. R. Elsawy, S. Lanteri, R. Duvigneau, J. A. Fan, and P. Genevet, “Numerical Optimization Methods for Metasurfaces,” Laser Photonics Rev. 14, 1900445 (2020).
    \bibitem{Molesky}
    S. Molesky, Z. Lin, A. Y. Piggott, W. Jin, J. Vuckovic, and A. W. Rodriguez, “Inverse design in nanophotonics,” Nat. Photonics 12, 659 (2018).
    \bibitem{Osher}
    S. Osher and J. A. Sethian, “Fronts propagating with curvature-dependent speed: Algorithms based on Hamilton-Jacobi formulations,” J. Comput. Phys. 79, 12–49 (1988).
    \bibitem{Preble}
    S. Preblea and M. Lipson, “Two-dimensional photonic crystals designed by evolutionary algorithms,” Appl. Phys. Lett. 86, 061111 (2005).
    \bibitem{Shokooh}
    M. Shokooh-Saremi and R. Magnusson, “Particle swarm optimization and its application to the design of diffraction grating filters,” Opt. Lett 32, 894-896 (2007).
    \bibitem{Wiecha}
    P. R. Wiecha, A. Arbouet, C. Girard, A. Lecestre, G. Larrieu, and V. Paillard, “Evolutionary multi-objective optimization of colour pixels based on dielectric nanoantennas,” Nat. Nanotech. 12, 163-169 (2016).
    \bibitem{Liu}
    C. Liu, S. A. Maier, and G. Li, “Genetic-Algorithm-Aided Meta-Atom Multiplication for Improved Absorption and Coloration in Nanophotonics,” ACS Photon. 7, 1716–1722 (2020).
    \bibitem{Piggott2020}
    A. Y. Piggott, E. Y. Ma, L. Su, G. H. Ahn, N. V. Sapra, D. Vercruysse, A. M. Netherton, A. S. P. Khope, J. E. Bowers, and J. Vučković, “Inverse-Designed Photonics for Semiconductor Foundries,” ACS Photon. 7, 569-575 (2020).
    \bibitem{Genevet}
    P. Genevet, F. Capasso, F. Aieta, M. Khorasaninejad, and R. Devlin, “Recent advances in planar optics: From plasmonic to dielectric metasurfaces,” Optica, 4, 139–152 (2017).
    \bibitem{Ding}
    F. Ding, A. Pors, S. I. Bozhevolnyi, “Gradient metasurfaces: a review of fundamentals and applications,” Rep. Prog. Phys. 81, 026401 (2018).
    \bibitem{Kamali}
    S. M. Kamali, E. Arbabi, A. Arbabi, and A. Faraon, Andrei, “A review of dielectric optical metasurfaces for wavefront control,” Nanophotonics 7, 1041-1068 (2018).
    \bibitem{GZheng}
    G. Zheng, H. Mühlenbernd, M. Kenney, G. Li, T. Zentgraf, and S. Zhang, “Metasurface holograms reaching 80\% efficiency,” Nature Nanotech. 10, 308-312 (2015).
    \bibitem{DLin}
    D. Lin, P. Fan, E. Hasman, and M. L. Brongersma, “Dielectric gradient metasurface optical elements,” Science 345, 298-302 (2014).
    \bibitem{Arbabi}
    A. Arbabi, E. Arbabi, Y. Horie, S. M. Kamali, and A. Faraon, “Planar metasurface retroreflector,” Nat. Photonics 11, 415 (2017).
    \bibitem{Khoram}
    D. Liu, Y. Tan, E. Khoram, and Z. Yu, “Training Deep Neural Networks for the Inverse Design of Nanophotonic Structures,” ACS Photonics 5,1365–1369 (2018).
    \bibitem{J.Mun}
    S. So, J. Mun, and J. Rho, “Simultaneous Inverse Design of Materials and Structures via Deep Learning: Demonstration of Dipole Resonance Engineering Using Core–Shell Nanoparticles,” ACS Appl. Mater. \& Interfaces 11, 24264–24268 (2019).
    \bibitem{L.Gao}
    L. Gao, X. Li, D. Liu, L. Wang, and Z. Yu, “A Bidirectional Deep Neural Network for Accurate Silicon Color Design,” Adv. Mater. 31, 1905467 (2019).
    \bibitem{F.Cheng_2019}
    W. Ma, F. Cheng, Y. Xu, Q. Wen, and Y. Liu, “Probabilistic Representation and Inverse Design of Metamaterials Based on a Deep Generative Model with Semi-Supervised Learning Strategy,” Adv. Mater. 31,1901111 (2019).
    \bibitem{X.Shi}
    X. Shi, T. Qiu, J. Wang, X. Zhao, and S. Qu, “Metasurface inverse design using machine learning approaches,” J. Phys. D: Appl. Phys. (2020).
    \bibitem{Goodfellow_2016}
    I. Goodfellow, Y. Bengio, and A. Courville, Deep Learning (MIT Press,2016).
    \bibitem{LeCun}
    Y. LeCun, Y. Bengio, and G. Hinton, “Deep learning,” Nature, 521, 436–444 (2015).
    \bibitem{Wiecha-review}
    P. R. Wiecha, A. Arbouet, C. Girard, and O. L. Muskens, “Deep learning in nano-photonics: inverse design and beyond,” Photonics Research 9, B182-B200 (2021).
    \bibitem{Malkiel}
    I. Malkiel, M. Mrejen, A. Nagler, U. Arieli, L. Wolf, and H. Suchowski, “Plasmonic nanostructure design and characterization via Deep Learning,” Light. Sci. Appl. 7, 60 (2018).
    \bibitem{Fowler}
    S. An, C. Fowler, B. Zheng, M. Y. Shalaginov, H. Tang, H. Li, L. Zhou, J. Ding, A. M. Agarwal, C. Rivero-Baleine, K. A. Richardson, T. Gu, J. Hu, and H. Zhang, “A Deep Learning Approach for Objective-Driven All-Dielectric Metasurface Design,” ACS Photonics 6, 3196–3207 (2019).
    \bibitem{Campbell_2020}
    Y. Li, Y. Wang, S. Qi, Q. Ren, L. Kang, S. D. Campbell, P. L. Werner, and D. H. Werner, “Predicting Scattering From Complex Nano-Structures via Deep Learning,” IEEE Access 8, 139983–139993 (2020).
    \bibitem{Zabaras}
    Y. Zhu, N. Zabaras, P.-S. Koutsourelakis, and P. Perdikaris, “Physics-constrained deep learning for high-dimensional surrogate modeling and uncertainty quantification without labeled data,” J. Comput. Phys. 394, 56–81 (2019).
    \bibitem{Chugh}
    T. Chugh, C. Sun, H. Wang, and Y. Jin, “Surrogate-assisted evolutionary optimization of large problems,” in High-Performance Simulation-Based Optimization, T. Bartz-Beielstein, B. Filipi c, P. Korošec, and E.-G. Talbi, eds. (Springer, 2020), pp. 165–187.
    \bibitem{COMSOL}
    "Introduction to {COMSOL} Multiphysics," COMSOL (2019).
    \bibitem{Zhelyeznyakov}
    M. V. Zhelyeznyakov, S. L. Brunton, and A. Majumdar, “Deep learning to accelerate Maxwell’s equations for inverse design of dielectric metasurfaces,” arXiv:2008.10632 [physics] (2020).
    \bibitem{B.Zheng}
    S. An, B. Zheng, H. Tang, M. Y. Shalaginov, L. Zhou, H. Li, T. Gu, J. Hu, C. Fowler, and H. Zhang, “Multifunctional Metasurface Design with a Generative Adversarial Network,” arXiv:1908.04851 [physics] (2020).
    \bibitem{J.Jiang}
    J. Jiang, D. Sell, S. Hoyer, J. Hickey, J. Yang, and J. A. Fan, “Free-Form Diffractive Metagrating Design Based on Generative Adversarial Networks,” ACS Nano 13, 8872–8878 (2019).
    \bibitem{P.Rodrigues}
    Z. Liu, D. Zhu, S. P. Rodrigues, K.-T. Lee, and W. Cai, “Generative Model for the Inverse Design of Metasurfaces,” Nano Lett. 18, 6570–6576 (2018).
    \bibitem{S.SoandJ.Rho}
    S. So and J. Rho, “Designing nanophotonic structures using conditional deep convolutional generative adversarial networks,” Nanophotonics 8, 1255–1261 (2019).
    \bibitem{A.Mall}
    A. Mall, A. Patil, D. Tamboli, A. Sethi, and A. Kumar, “Fast design of plasmonic metasurfaces enabled by deep learning,” J. Phys. D: Appl. Phys. 53, 49LT01 (2020).
    \bibitem{Goodfellow_2020}
    I. Goodfellow, J. Pouget-Abadie, M. Mirza, B. Xu, D. Warde-Farley, S. Ozair, A. Courville, Y. Bengio, “Generative adversarial networks,” Communications of the ACM 63, 139-144 (2020).
    \bibitem{ResNet}
    C. Szegedy, S. Ioffe, V. Vanhoucke, and A. Alemi, “Inception-v4, Inception-ResNet and the Impact of Residual Connections on Learning,” arXiv:1602.07261 [cs] (2016).
     \bibitem{Adibi}
    M. Zandehshahvar, Y. Kiarashi, M. Chen, R. Barton, and A. Adibi, “Inverse design of photonic nanostructures using dimensionality reduction: reducing the computational complexity,” Opt. Lett. 46, 2634-2637 (2021).
\end{thebibliography}
\end{document}